%% file: ckpt_modeling.tex
\documentclass[times, 12pt, onecolumn]{article}
\usepackage{amssymb,amsmath,shortvrb,graphicx,psfrag,subfigure,alltt,url}
\usepackage[ruled,algonl]{algorithm2e}

\renewcommand\baselinestretch{0.92}

\newcommand{\refeq}[1]{Equation.~(\ref{#1})}

\newcommand{\refsec}[1]{Section~\ref{#1}}
\newcommand{\reffig}[1]{Fig.~\ref{#1}}

\newcommand{\reftab}[1]{Table~(\ref{#1})}

\begin{document}
\sloppypar

\title{An Adaptive Checkpointing Scheme for Peer-to-Peer Based Volunteer Computing Work Flows} 

\author{Lei Ni, Aaron Harwood \\
NICTA Victoria Labs,  Australia, \\
Department of Computer Science and Software Engineering, \\
The University of Melbourne, Australia \\
{http://www.cs.mu.oz.au/p2p}}

\maketitle 
\begin{abstract} 
Volunteer Computing, sometimes called Public Resource Computing, is an emerging
computational model that is very suitable for work-pooled parallel processing.
As more complex grid applications make use of work flows in their design and
deployment it is reasonable to consider the impact of work flow deployment over
a Volunteer Computing infrastructure. In this case, the inter work flow I/O can
lead to a significant increase in I/O demands at the work pool server. A
possible solution is the use of a Peer-to-Peer based parallel computing
architecture to off-load this I/O demand to the workers; where the workers can
fulfill some aspects of work flow coordination and I/O checking, etc.  However,
achieving robustness in such a large scale system is a challenging hurdle
towards the decentralized execution of work flows and general parallel
processes. To increase robustness, we propose and show the merits of using an
adaptive checkpoint scheme that efficiently checkpoints the status of the
parallel processes according to the estimation of relevant network and peer
parameters. Our scheme uses statistical data observed during runtime to
dynamically make checkpoint decisions in a completely decentralized manner. The
results of simulation show support for our proposed approach in terms of
reduced required runtime.     
\end{abstract} 

\section{Introduction}
The emerging concept of Volunteer Computing~\cite{935629, volunteercomputing}
has proved its very large capacity for high performance computing in a series
of highly successful projects, namely SETI@Home~\cite{setiathome},
Folding@Home~\cite{larson-2003} and etc. For example, in SETI@Home, more than
1.5 million desktops world wide are actively contributing their processors for scientific
computing and that led to over 250 TeraFLOPS of processing power as of
2007\footnote{{\tt http://boincstats.com/stats/project\_graph.php?pr=sah},
accessed on 27th, Sept. 2007}. This is comparable to the current fastest computer, 
IBM Blue Gene, which is offering sustained speeds of 360 TeraFLOPS.

\subsection{Work Pooling and Work Flows}

The Berkeley Open Infrastructure for Network Computing (BOINC)~\cite{anderson04} is a
generalization of SETI@Home and others which uses a work pool model, where a
work pool server coordinates the work being done over a number of workers.
Workers pull work units and push results. 

\begin{figure}
\centering
\subfigure[Work flow coordination in a work pool model. Inter-work flow communication takes place via the server.]{\label{workflow}\includegraphics[width=6cm]{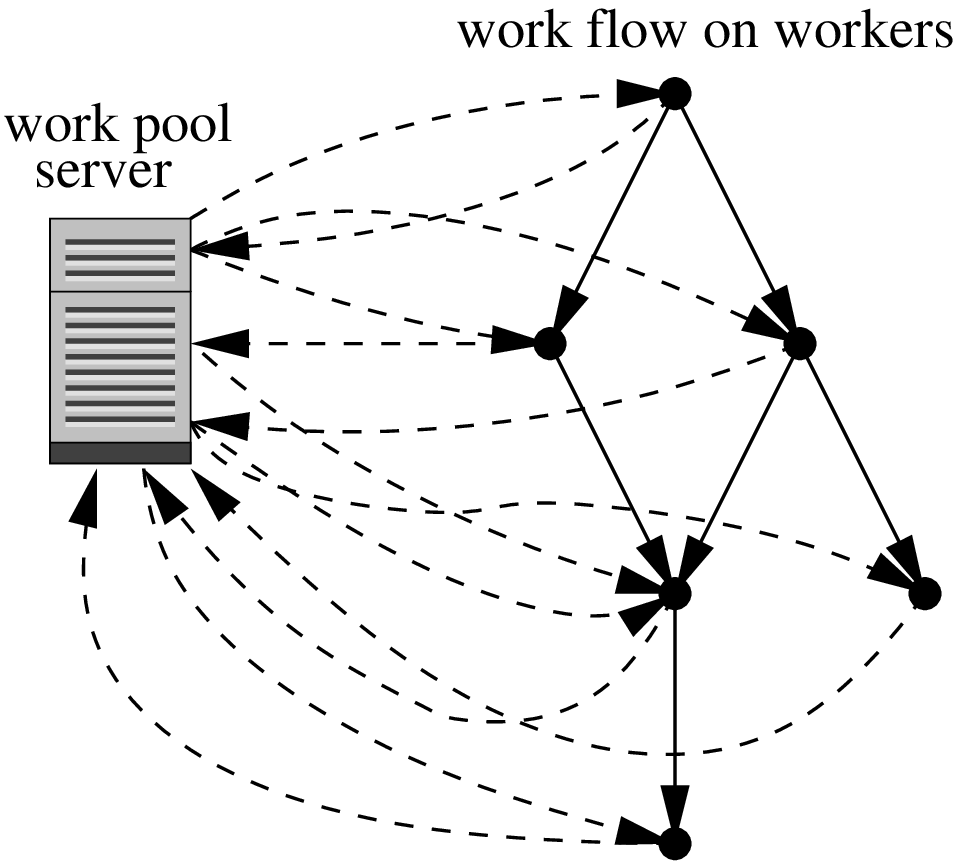}}
\hspace{1cm}
\subfigure[Work flow coordination in a work pool model with peer-to-peer
networking among the workers.  Inter-work flow communication takes place via
the peer-to-peer network.]{\label{workflowp2p}\includegraphics[width=6cm]{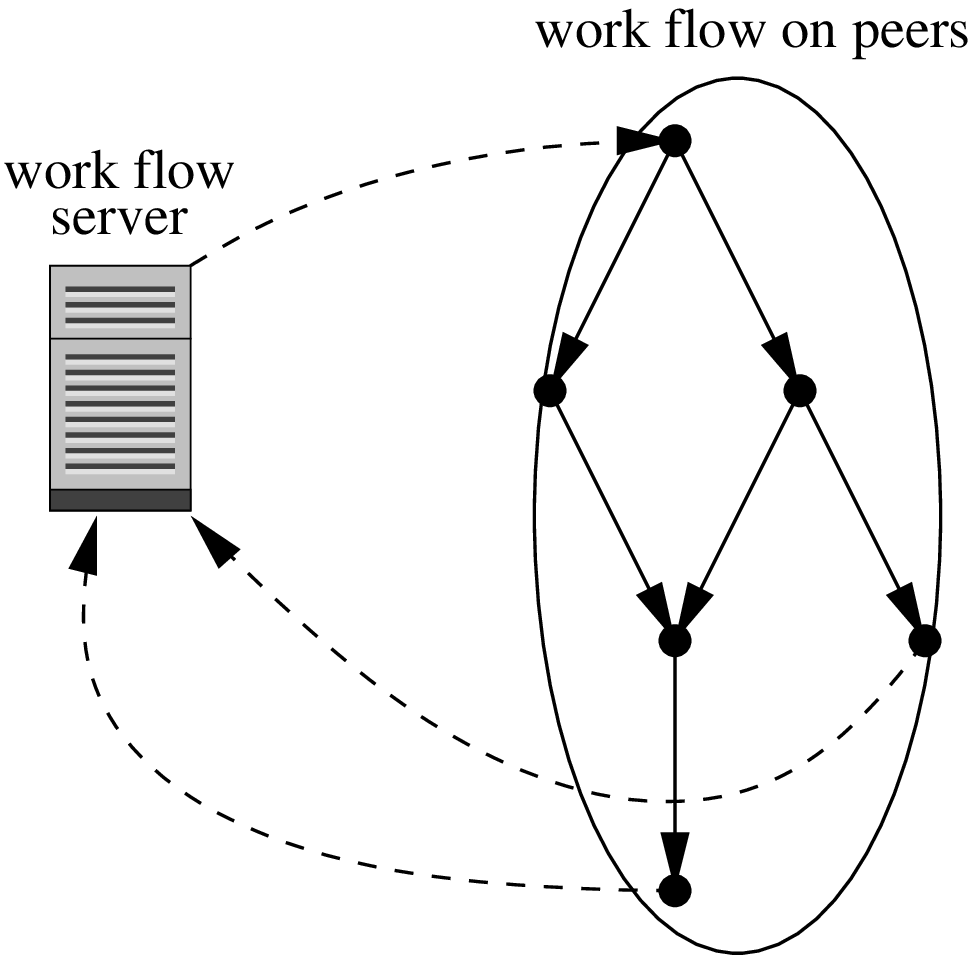}}
\caption{Work pooling and work flows.}
\label{workflowfig}
\end{figure}

The work pool model can be readily extended to handle work flows and doing so
will support work flow based grid applications. \reffig{workflow} shows a work
pool server that is coordinating the flow of work among several workers (solid
circles). Each step of the work flow (shown as solid arrows) requires
communication to and from the server (dashed arrows). The communication is
required for multiple reasons: (i) the workers usually cannot communicate
directly with each other, (ii) to avoid malicious volunteers the results from
workers need to be scrutinized for correctness and (iii) a large work flow
requires checkpointing to be efficient.

Clearly, the server communication requirements increase proportionally to the
number of work flow steps. More so, if the work flow contains iterative
elements, i.e. cycles, then the communication to the server will increase
proportional to the complexity of the iterations. Such communication could
quickly slowdown more complicated work flows.

In general, one can model the work flow as a parallel process, i.e. as a
message passing parallel program. A simple work flow is like a pipeline of
tasks in a parallel process. It is also particularly suitable to do so for the
cases when the work flow includes cycles with large numbers of iterations.  In
order to eliminate the communication to the server for each work flow step we
propose the use of a peer-to-peer based parallel architecture that allows the
workers to scrutinize and checkpoint their results independently of the server.
In this case, shown in \reffig{workflowp2p}, intra-work flow communication is
done using a peer-to-peer network, and only inter-work flow communication
requires the server.



\subsection{P2P Based Parallel Computing}
 
P2P based parallel processing systems~\cite{mpichopen, dvm, p2pmpi} have already been built in the past
few years and the previous studies~\cite{dvm} have shown that with proper designs the P2P based approach can 
potentially be used to collect free idle CPU cycles from the Internet efficiently with reduced maintenance cost
and reduced risk of single point failure. In such P2P platforms, nodes are indexed in a Distributed Hash Table 
(DHT) overlay~\cite{chord, mspastry, kademlia} and messages for parallel computing are routed in multiple hops 
and in a decentralized way. This decentralized design enables message passing programs on such P2P platforms and 
allows for running a wide range of existing message passing programs. 

\subsubsection{Node Failure}
For both the above mentioned new and traditional architectures of Volunteer Computing, one of the common issues 
is how to handle node failure and departure. In the context of this paper, as both the node failure and departure
would immediately make the computing and storage resources of the node unavailable, we choose to use the term 
failure to represent such events. Different from other parallel processing domains where computing nodes are 
usually dedicated and well maintained, the nodes in a Volunteer Computing platform can disconnect at any time 
and this can happen relatively frequently. The traditional Volunteer Computing projects like SETI@Home and Folding@Home have a 
simple `deadline' approach where each work unit is assigned together with a deadline for reporting results 
and thus the system can reassign the work unit again to another node if the results can't be reported on time due 
to failure or machine departure. This approach is fine with data and parametric parallel programs as in such 
programs each work unit is usually independent of each other and can be recomputed by any other node at 
any time. However this mechanism is not sufficient to support parallel processing which use message passing.

\subsubsection{Checkpointing}

One of the long applied methods to counter nodes failure and departure in message passing systems is to 
employ checkpoint and restart~\cite{DBLP:journals/csur/ElnozahyAWJ02,ChandyL85} in which the status of the 
message passing job is saved regularly and stored on a reliable storage so that the progress can be rolled back 
to the last known status when any involved node fails (e.g. disconnect from the network or crash due to hardware 
or software failure). For example, the Chandy-Lamport algorithm~\cite{ChandyL85} is used in our P2P DHT based 
P2P-DVM system~\cite{dvm} to capture the status of the MPI job every $T$ seconds where the fixed number $T$ is 
a parameter for each job chosen by the user, the captured processes status are saved on a P2P based distributed
storage system. The experiments have shown such checkpoint-rollback protocol causes little overhead for fault free 
running compared to other approaches and the jobs can be finished eventually with the presence of peer failure events 
as long as the peer failure events (peer disconnection or machine crash) can be detected and the value of $T$ is
chosen carefully. A similar approach is also found on a few cluster and networked workstation based systems~\cite{BHKLC06}. 

The above mentioned checkpoint-rollback approach has demonstrated it usefulness in our previous work but it 
also introduced the question of how to choose the value of the checkpoint interval $T$ and is there any 
optimal value of $T$ for a given system at the given time so that the overall performance of the system can be 
improved? The existing implementation requires users to manually choose the checkpoint interval before the job is 
submitted which is quite difficult for 
users without adequate knowledge about the running environment (e.g. available bandwidth, failure rate, etc.). 
A method does not take the dynamism of the P2P networks into consideration will not adapt 
well when the nodes failure rate or available bandwidth change from time to time. Such a non-optimized approach
may also create another major overhead for the overall performance as some resources need to be consumed when storing
the process context and store it on the network. We aim to address the issue in this work by proposing an 
adaptive checkpoint scheme which can automatically adjust the checkpoint interval during runtime to reduce the 
total overheads introduced due to checkpointing and restarting.    

\subsection{Contribution}
In this work, we propose an adaptive scheme for automatically making checkpoint decisions during runtime, 
the decision method dynamically decides the optimized checkpoint interval based on the estimated network 
conditions which in turn is based on the statistical data collected online. The scheme is designed to be
completely decentralized. We use simulations to show when our model is better than the fixed checkpoint 
interval in terms of reduced total runtime.   

\subsection{Related Works}
The idea of optimizing system performance using estimated P2P network conditions has been applied in
a few projects at different levels of the P2P software stacks. In ~\cite{GhinitaT06} and ~\cite{mspastry} 
the idea of probability based stabilization is proposed to better control the cost of stabilization based 
on the estimation of both the P2P network size and peer failure rate and this will improve P2P routing 
success rate. Similar idea is applied in the gossip protocols~\cite{gossip} so with the estimated P2P 
network size, the protocol will be able to compute the number of gossip targets to reach for gossip messages. 
However, to the best of our knowledge, we are the first to propose and apply such method for P2P based 
parallel computing.

A novel peer availability prediction algorithm using multiple 
predictors was proposed in~\cite{predictor} and it demonstrates good performance for possible distributed 
applications. In that work, each peer's availability is predicted based on its historical connection 
and disconnection statistics. Applications can benefit from such availability prediction 
in their application level planning. This algorithm depends on the availability of the log data which may not be available
for some peers, e.g. peers which just have the software installed and thus don't have their log data
at all. For example, it has been reported that SETI@Home~\cite{setiathome} attracts around 2,000 new 
machines daily~\cite{volunteercomputing}, it is clear that such new peers will not be able to predict 
their availability by using~\cite{predictor}. E.g. two weeks logging data is used for training the 
predictor in~\cite{predictor}. The computational grid community applied a similar approach~\cite{NurmiBW05} 
to predict the available of the grid computing nodes based on their uptime log, this is acceptable for 
the grid systems where nodes are all professionally managed. This approach will suffer in a P2P network 
for the same reason mentioned above. 

\section{Typical Peer Uptimes and Checkpointing Impact}
\label{limitations}
On traditional dedicated parallel processing platforms like clusters, the machine failure can be 
regarded as exceptional events. For example, as of the time of writing, the PRAGMA grid web portal 
shows its 207 hosts have an average uptime of 68.5 days\footnote{{\tt http://goc.pragma-grid.net/cgi-bin/scmsweb/uptime.cgi}, 
accessed on 21 June, 2007}. Given such average uptime of more than two months, the risk of software and hardware 
failures can be mitigated with the help of the checkpoint and restart. The fixed checkpoint 
interval can work well with the interval set to be approximately checkpointing a few times daily as the 
probability for each job, which may run up to a few days, to have multiple failures during its life time 
is quite low and thus the overhead that can be introduced by the fixed checkpoint interval is not significant. 

However the P2P network conditions present a new challenge for running message passing jobs on P2P networks. 
In general, the P2P networks consist of a large number of peers which are Internet users' desktop and laptop 
PCs, these machines can disconnect from the P2P networks quite frequently and we found the rate of such 
departure, which will cause the parallel processing on the peer to fail immediately, dominate the reasons 
of job failure and it varies widely from time to time. 

In order to highlight the running environment of parallel processing systems over P2P networks, we discuss
traces data from three globally deployed P2P networks (There is currently no such peer failure statistics 
available from Volunteer Computing systems like SETI@Home.) 
The peer lifespan trace project\footnote{{\tt http://www.aqualab.cs.northwestern.edu/projects/lifeTrace.html}} 
monitored over 500,000 peer sessions in the Gnutella file-sharing P2P network for a week, the trace 
data shows the average peer session time is about 121 minutes, that is the computing nodes will become 
unavailable in just two hours on average. Another real P2P system called Overnet is measured for 7 days 
in~\cite{BhagwanSV03} and the results indicated a similar average session time of 134 minutes for its 
1468 peers observed. In another more recent study, a detailed measurement study of the popular Bittorrent 
P2P network is provided by~\cite{Powlese05BitTorrent} and the active probe part of the 
\emph{Delft Bittorrent Dataset}\footnote{\tt {http://multiprobe.ewi.tudelft.nl/dataset.html}} was obtained by 
continuously measuring more than 180,000 Bittorrent peers and the average session time is 104 minutes which 
is consistent with the results found in both the Gnutella and the Overnet trace data. Such short average
session time suggests that robustness design is essential as even a typical short job which takes a few 
hours will suffer from a few failures. 

\begin{figure*}
\centering
\subfigure[The peer failure in Gnutella can loosely fits the exponential distribution.]{\label{gnutella}\includegraphics[width=8cm]{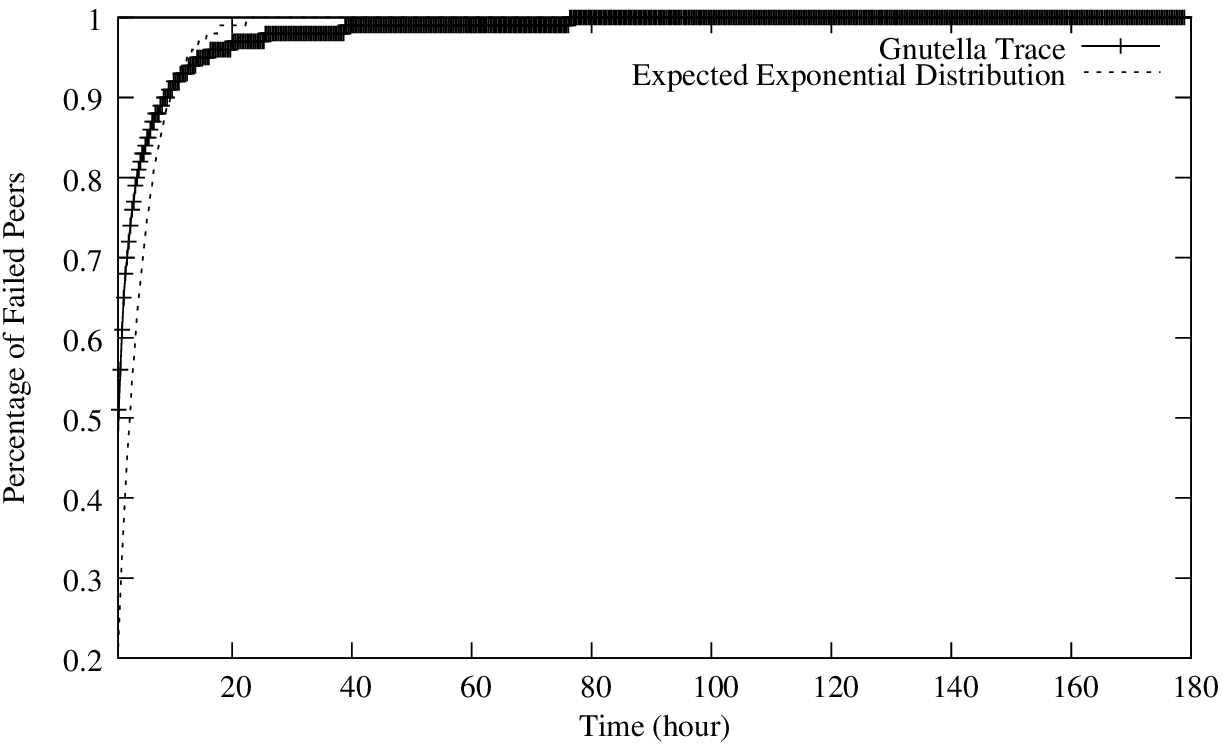}}
\subfigure[The short term failure rate in Overnet is highly dynamic.]{\label{overnet}\includegraphics[width=8cm]{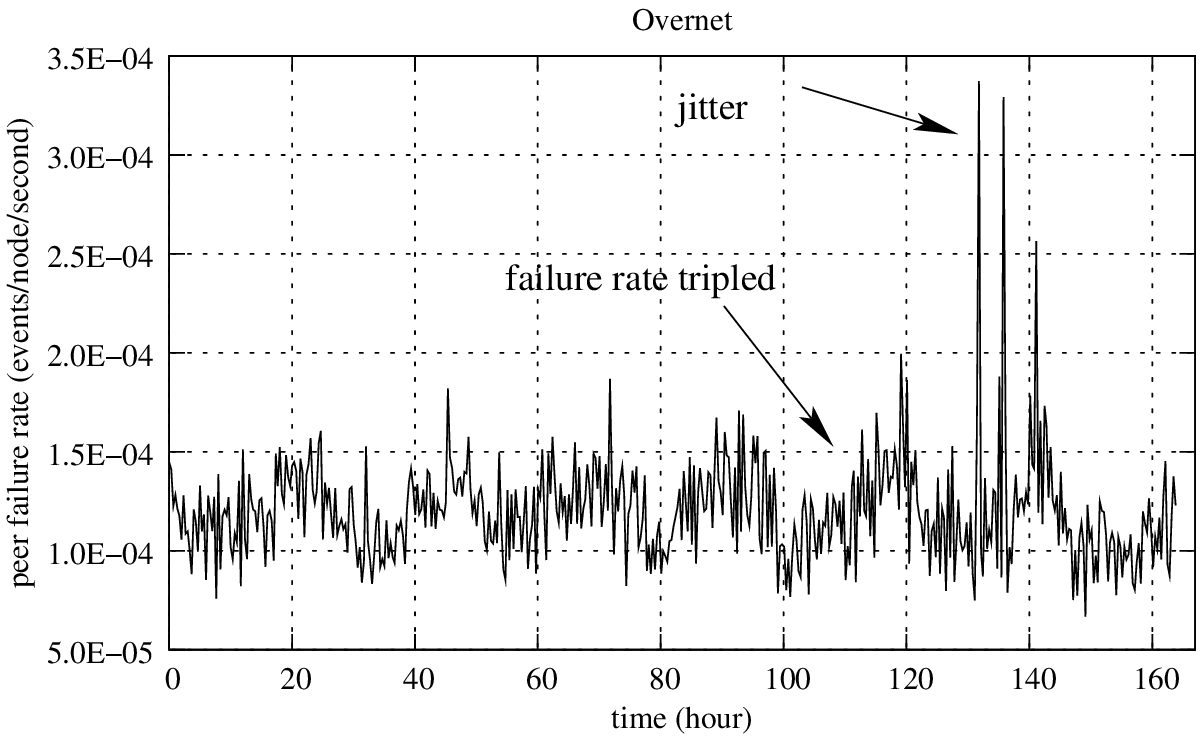}}
\caption{The Peer Failure in the Gnutella and Overnet P2P Networks.}
\label{failurerate}
\end{figure*}

As visualized in \reffig{gnutella}, the Gnutella trace data clearly show that most of peers 
will leave the network in just several hours and the failure rate curve can loosely fit the 
expected exponential distribution curve which has the mean time before failure parameter 
equals to the Gnutella's average peer life time. 
To enable the system to eventually produce the correct result within reasonable runtimes in such a
P2P environment, we consider automatically adapting to the current network conditions should improve upon
a fixed interval in~\cite{dvm}, because:
\begin{itemize}
\item Too small checkpoint interval $T$ would lead to many unused checkpoint operation and total 
fault free running overhead in terms of both required job execution time and bandwidth consumption for
uploading checkpoint image files. 
\item Large $T$ reduces overheads mentioned above but can increase the impact of restarting, as the
runtime wasted by each restart has the upper boundary of $T$.
\item Different from the traditional platforms (e.g. clusters) where the system mean time before failure 
(MTBF) can be estimated offline and will usually stay unchanged during the job, the average MTBF value of 
peers in the P2P network are dominated by nodes' failure events and such rate can change from time to time. 
The fixed checkpoint interval $T$ may have sub-optimal performance in real deployed Internet environment as it 
does not adapt to such changes. 
\end{itemize}

\section{Adaptive Checkpoint Scheme}
\label{scheme}
In our approach we analyse the efficiency of our checkpoint decision scheme which tries to optimize the 
utilization of total runtime. In this section, we define the system parameters first and then describe 
how some of them are estimated in \refsec{estimation} and present our model in \refsec{model}.

The parameters used in this work are summarized in \reftab{parameters}.
\begin{table*}[htbp]
\begin{center}
\caption{The Parameters Used in the Adaptive Checkpoint Scheme.}
\label{parameters}
\begin{tabular}{|c|c|p{0.5\linewidth}|}
        \hline
Name                 &  Symbol    & Definition \\
        \hline
Peer Failure Rate    &  $\mu$     & Peer failure is dominated by the peer departure events and we model \\
                     &            & it as exponential distribution~\cite{tianjing07, GhinitaT06} with the rate parameter $\mu$.  \\
	\hline
The number of peers  &  $k$       & The number of peers used in the job. \\
        \hline
Checkpoint rate      &  $\lambda$ & How often the status of the job is checkpointed, \\ 
                     &            & the checkpoint interval is thus $\frac{1}{\lambda}$.  \\
	\hline
Checkpoint overhead  &  $V$	  & Extra runtime caused by the checkpoint in terms of runtime.  \\ 
        \hline
Wasted computation time & $T_{wc}$ & Unsaved computation progress to be lost on failure. \\
        \hline
Image download overhead & $T_d$    & The required amount of time to download the checkpoint image.\\ 
	\hline
\end{tabular} 
\end{center}

\end{table*}

We model the peer failure in this work as exponential distribution as this has
been suggested in some previous works~\cite{tianjing07, GhinitaT06} and accepted
by the community. Tian and Dai ~\cite{tianjing07} have recently reported that once 
peers are grouped into different categories according to their average life time (e.g. 
long, medium and short life time), peers' failure can be even better fitted to the 
exponential distribution. 

\begin{figure}
\psfrag{1/lambda}{$1/\lambda$}
\psfrag{V}{$V$}
\psfrag{Twc}{$T_{wc}$}
\psfrag{td}{$T_d$}
\psfrag{c1}{$c_1$}
\psfrag{c2}{$c_2$}
\psfrag{c3}{$c_3$}
\centering{\includegraphics[width=0.65\linewidth]{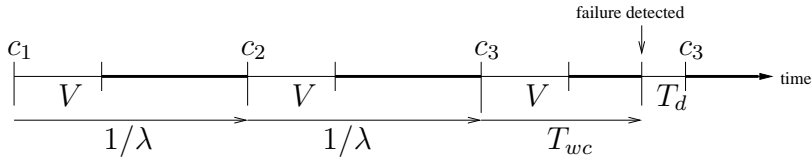}}
\caption{The relationship of the parameters with an example failure.}
\label{timeline}
\end{figure}

During the execution of a message passing job, the time line of the possible
events and the relationship of the above parameters are explained in
\reffig{timeline}. We compute an optimized value of $\lambda$ for the P2P
network conditions according to the observation and estimation of the above
$\mu$, $V$ and $T_d$ parameters.

\subsection{Online Parameter Estimation}
\label{estimation}
The estimation of peer failure rate $\mu$, the checkpoint overhead $V$ and the image download overhead 
$T_d$ which we use to compute the optimized value of $\lambda$ is given below. The scheme used in our 
P2P based system must be decentralized to ensure the scalability, each peer would be desirable to carry 
out estimations based on its own observed information about the network conditions, any centralized 
monitoring component should be strictly avoided due to both the availability and scalability concerns. 
We also make sure that all the network estimation employed here will not increase the communication 
complexity, in fact the number of messages exchanged for making the checkpoint decisions in this proposed 
model is the same as our previous naive fixed interval approach used in~\cite{dvm,mpichopen}.

\subsubsection{Peer Failure Rate $\mu$}
\label{failureestimation}
Given the goal of this study is to come up with an optimized scheme which can automatically scale the
checkpoint interval for different P2P network conditions, an accurate estimation of the value $\mu$ 
is important. We did a detailed study~\cite{estimation} to find the comparative performance for the possible failure rate 
estimation methods and the results of our study indicate the Maximum Likelihood based method~\cite{estimation} out performs 
the commonly used estimation methods in most cases. 

In this work we use the Maximum Likelihood method for estimating peer failure rate, where 
$\mu$ is estimated as: 
\begin{equation}
\mu=\frac{1}{\;\;\overline{t_{l}}\;\;}=\frac{K}{\sum_{i=0}^K{t_{l,i}}}
\label{max-likelihood}
\end{equation}
where $\overline{t_{l}}$ is the average life time observed by the peer while $K$ is the required number of 
observed failures to compute a new estimated failure rate. We use a simple but effective cooperative scheme 
for collecting peer failure observations. Each peer shares its failure observation with its neighbours, and
their neighbours, thus effectively allows each peer to monitor the status of its neighbouring peers and the 
neighbours of its neighbours.    

\subsubsection{Checkpoint Overhead $V$}

The checkpoint overhead means how much time each checkpoint operation
will slow down the parallel process. The overhead is caused by (\emph{i}) dumping
the whole memory space used by the process will have some stress on the memory bandwidth, (\emph{ii}) 
compressing the checkpointed status costs some processing cycles (\emph{iii}) the available bandwidth 
need to be consumed to upload the checkpoint image file to a reliable storage and this will slow down 
the message passing for computing purpose. We perform the estimation online and do not rely on the historical 
data of the required runtime without checkpoint as the P2P network environment can be highly dynamic and 
the peers used for each run can be totally different, it is also not practical to expect a long running job 
to finish without any peer failure during execution. We estimate 
the checkpoint overhead $V$ by observing both the average CPU usage of the parallel application process and 
the number of messages exchanged for computing purpose. To estimate the value of $V$, we first run the 
parallel application without checkpoint for $t$ minutes and the average CPU usage $P_1$ and the totally
number of incoming and outgoing messages $M_1$ will be recorded. We then switch on the checkpoint with a 
relative small interval and run for another $t$ minutes. If the number of checkpoint performed is $y$, 
the average CPU usage and the number of exchanged messages on the local peer are $P_2$ and $M_2$ respectively,
then we estimate two separate $V$ based on both the CPU usage and network IO statistics as:
\begin{equation} 
V={\frac{\left(P_1-P_2\right)\left(M_1-M_2\right) t}{2 P_1 M_1 y}}
\end{equation}

\subsubsection{Checkpoint Image Download Time $T_d$}

After the job is submitted and we have the estimated value
of $V$, we set $T_d$ to be same as $V$ as its initial value. When the first checkpoint image is captured
and uploaded using the estimated $\lambda$, we download the checkpoint image from the P2P network while
still keeping the parallel processing running at the background and the download time becomes a more 
accurate estimated value of $T_d$. If any restart is performed during the execution then the download
time will be used as the $T_d$. This is also due to the desire that we want to predict the optimized value
of $\lambda$ based on the recent network conditions. 

\subsubsection{Global Versus Local Estimation}
The above $\mu$, $V$ and $T_d$ are computed according to the local knowledge
on each peer and thus they reflect the local peer's estimation. It can be more accurate if these local
estimations can be combined, for example take the average value, to form better global estimations. For example, 
as the coordinated global checkpoint~\cite{ChandyL85} is used in our system in which all involved peers will 
checkpoint the status of the job once any peer issue the checkpoint command, if every peer issued such 
command according to their results of $\lambda$ which in turn depends on their own estimation of $\mu$, then
the global checkpoint rate for the system would be decided by the smallest estimation of $\mu$ on all peers. 
In this case, an average $\mu$ based on at least a few peers' estimation 
can be more accurate. We make each peer to periodically piggyback its own most recent 
estimation of $\mu$, $V$ and $T_d$ in its messages used for parallel computing message passing which are sent 
to other peers. On receiving such piggybacked values, the peer can use them to calculate the global estimated 
$\mu$, $V$ and $T_d$ using the average of these estimated values by different peers. This approach doesn't 
increase the communication cost a lot because no extra message is require, only a few messages' length will be 
slightly increased to carry and exchange the attached locally estimated values. 

\subsection{Runtime Utilization Based Adaptive Scheme}
\label{model}
In our scheme, we define the notion of \emph{average cycle utilization}, $U$,
which is the fraction of time in a cycle time, $1/\lambda$, that is spent doing
useful computation. The remaining time in the cycle is overheads introduced by
the checkpointing and costs of restarting.  In basic terms:
\begin{equation}
U =\frac{\frac{1}{\lambda}-C}{\frac{1}{\lambda}}=1-\lambda\, C
\label{u-eq}
\end{equation}
where $C$ is the average overhead and failure costs per cycle. We use $U$ as an
approximate utilization of the system. In this simple case, if the overheads
exceed the cycle time then the system operates at zero percent efficiency. The
value for $C$ is computed from a predicted failure rate and the various
overheads.  To explain the model that we use for the impact of failure we first
consider a simplified model in which only a single peer is involved. 

\subsubsection{Single Peer Model}

In this setting, the peer has the failure rate of $\mu$ and it uses
checkpointing and restarting with the costs of $V$, $T_{wc}$ and $T_d$. Because
the peer failure is modeled as exponential distribution~\cite{tianjing07,
GhinitaT06} the probability of the job (i.e. peer) to fail at time $t$ is:
\begin{equation}
P_{f}(t)=\mu e^{- \mu t}
\label{eq-pf}
\end{equation}
which is the probability density function of the exponential distribution. As the computation result between the 
last checkpoint and the failure will be wasted when the job has to roll back to the last checkpoint for restart, 
the expected wasted computation time for each failure is given by:
\begin{equation}
T_{wc}=\sum_{i=0}^\infty{
\int_{\frac{i}{\lambda}}^{\frac{i+1}{\lambda}}{P_f(t) \left(t-\frac{i}{\lambda}\right)} dt}
=\frac{1}{\mu}-\frac{1}{e^{\frac{\mu}{\lambda}}-1}\, \frac{1}{\lambda}=\frac{1}{\mu} - \overline{c}\frac{1}{\lambda}
\end{equation}
%
%
where $\overline{c}$ is the average fault free running cycles for each expected
failure, i.e. how many checkpoint operations are expected to succeed on average
before every failure. An alternative way to derive $\overline{c}$ is:
\begin{equation}
\overline{c}=\sum_{i=0}^\infty{i \int_{\frac{i}{\lambda}}^{\frac{i+1}{\lambda}}{P_f(t)}\;dt}=
\frac{1}{e^{\frac{\mu}{\lambda}}-1}
\label{cbar}
\end{equation}
%
%
%
%
%

\subsubsection{Multiple Peer Model} 

In a multi-peer environment, $k$ peers work together on the same job and they do checkpoint and restart 
in a coordinated manner according to the Chandy-Lamport algorithm~\cite{ChandyL85} to count the possible failures. 
In this setting, the job will fail and a restart must be issued once any one of the $k$ peer fails. Since the 
failure rate for each job is $\mu$ then the $k\,\mu$ would be the failure rate for the job. The above \refeq{eq-pf} 
to \refeq{cbar} can be trivially modified to describe the multi-peer environment:

\begin{equation}
P_{f}'(t)=k \mu e^{- \mu k t}
\label{eq-pfm}
\end{equation}
  
\begin{equation}
T_{wc}'=\sum_{i=0}^\infty{
\int_{\frac{i}{\lambda}}^{\frac{i+1}{\lambda}}{P_f'(t) \left(t-\frac{i}{\lambda}\right)} dt} 
=\frac{1}{k \mu}-\frac{1}{e^{\frac{\mu k}{\lambda}}-1} \frac{1}{\lambda} \\
\end{equation}

Similarly, $\overline{c}'=\frac{1}{e^{\frac{k\,\mu}{\lambda}}-1}$.

\subsubsection{Average Cycle Utilization}

We formulate the average cycle utilization from the total overheads: 
\begin{equation}
C=V+\frac{\left(T_{wc}'+T_d\right)}{\overline{c}'}
\end{equation}
\begin{equation}
U=\begin{cases}
1-C\,\lambda & \text{if $C\,\lambda<1$ },\\
0& \text{if $C\,\lambda>1$.}
\end{cases}
\label{eq-u}
\end{equation}
The desired $\lambda$ leads $U$ to peak when $\dfrac{dU}{d\lambda}=0$ and thus can be computed as:
\begin{equation*}
\lambda=\frac{k \mu}{lamberW\left[\left(v k \mu- T_d k \mu - 1\right)\left(T_d k \mu+1\right)^{-1} e^{-1}\right]+1},
\end{equation*}
%
%
%
where the above $lambertW\left(\right)$ function is the Lambert W function. 

\refeq{eq-u} can be used to judge whether the number of peers used
for the parallel processing job is reasonable (i.e. the job can at least
progress) given the current network conditions. According to the definition of
the \emph{utilization}, a positive $U$ means the job is at least still
progressing while a zero value means that no significant progress for the parallel process
can be made. When the estimated $\mu$, $V$, $T_d$ and the above optimal
$\lambda$ together with a user specified job parameter $k$ lead to $U=0$ in
\refeq{eq-u}, it suggests the number of peers used for the job is too large. 

\input{evaluation}

\section{Conclusion and Future Work}
The role of Peer-to-Peer based parallel computing in future grid systems is
becoming more necessary as we consider large scale Volunteer Computing that
employs work flow based grid applications.  Given a highly dynamic environment
like current P2P networks where most peers usually just keep online for hours,
and other network conditions are also changing, it is complicated and
inefficient for the users to decide how to choose suitable system parameters
like the checkpoint interval. In this work we have proposed and evaluated an
adaptive checkpoint scheme for parallel processing systems over Peer-to-Peer
networks. The evaluation results show that our proposed approach which is based
on network condition estimation is almost always better than the naive fixed
checkpoint interval approach in terms of reduced runtime. The results of our
work presented in this paper can potentially be used in the next generation of
Peer-to-Peer based parallel processing systems to effectively handle churn
events and provide robustness for the system. As a part of the future work, we
are going to test this scheme in real deployed P2P message passing systems over
the Internet. It would also be an interesting topic to study both the
possibility and the feasibility of combining the historical log and the real
time network conditions observation data to predict with higher accuracy of the
parameters of the running environment.

\bibliographystyle{abbrv}
\bibliography{ckpt_modeling}

\end{document}

%% file: evaluation.tex
\section{Simulation Evaluation}
\subsection{Setup for the Evaluation}

To evaluate our proposed adaptive checkpointing scheme we provide performance comparison using simulation. 
We extended the P2P simulator used in~\cite{estimation} to simulate the running of P2P based message passing 
programs under the affect of peer failure events. In this simulator a typical P2P network is simulated where 
peers can connect and disconnect according to exponential distribution, each peer knows several other peers 
(neighbors) and peer departure events can be detected during each peer's stabilization~\cite{chord, mspastry}. 
The observed failure events can be used for network failure rate estimation as described in \refsec{failureestimation}. 
Message passing jobs can be simulated by specifying the number of peers to use and its required runtime in a 
fault free environment. When the job is submitted, a list of peers is chosen at random to simulate the 
message passing job and the progress of such jobs can be saved periodically according to either fixed checkpoint 
interval or dynamically picked intervals produced by our adaptive scheme. The status of the job will always be 
rolled back to its previous saved checkpoint upon peer failure events. The peer failure rate, overheads, required 
runtime for the job in a fault free environment and the fixed checkpoint interval used in the naive fixed 
checkpoint interval approach can be set as the parameters of the simulation. 

We compare the performance between the naive fixed checkpoint interval approach
and our adaptive checkpointing scheme with different network conditions,
including different peer departure rates, various available bandwidth and types
of parallel processing programs. We choose to do the evaluation in such a
simulated environment instead of in real deployed systems because: \emph{(i)}
to the best of our knowledge, there is currently no widely deployed P2P
parallel processing systems that supports both message passing and
checkpointing and \emph{(ii)} the simulated environment allows us to test our
proposed scheme in different network conditions which is essential to this
work. 

The metric called \emph{relative runtime} is used intensively in this section to compare the performance of our 
proposed approach and the naive fixed checkpoint interval approach. The relative runtime means the percentage of 
runtime using a fixed checkpoint interval compared to the runtime required by our adaptive scheme. 

\begin{equation}
Relative Runtime = \frac{Runtime\ using\ fixed\ checkpoint\ interval}{Runtime\ using\ the\ adaptive\ interval} \times 100\%
\end{equation}

The proposed adaptive checkpoint scheme would outperform the fixed checkpoint interval for a given checkpoint rate if the relative 
runtime is larger than 100\%. The MTBF is used to represent the departure rate in this section for better readability 
and the departure rate is $\frac{1}{MTBF}$ as the failure events are of exponential distribution.   

\subsection{Evaluation}
\begin{figure*}
\begin{minipage}[b]{0.46\linewidth}
\centering
\includegraphics[width=6.0cm, angle=270]{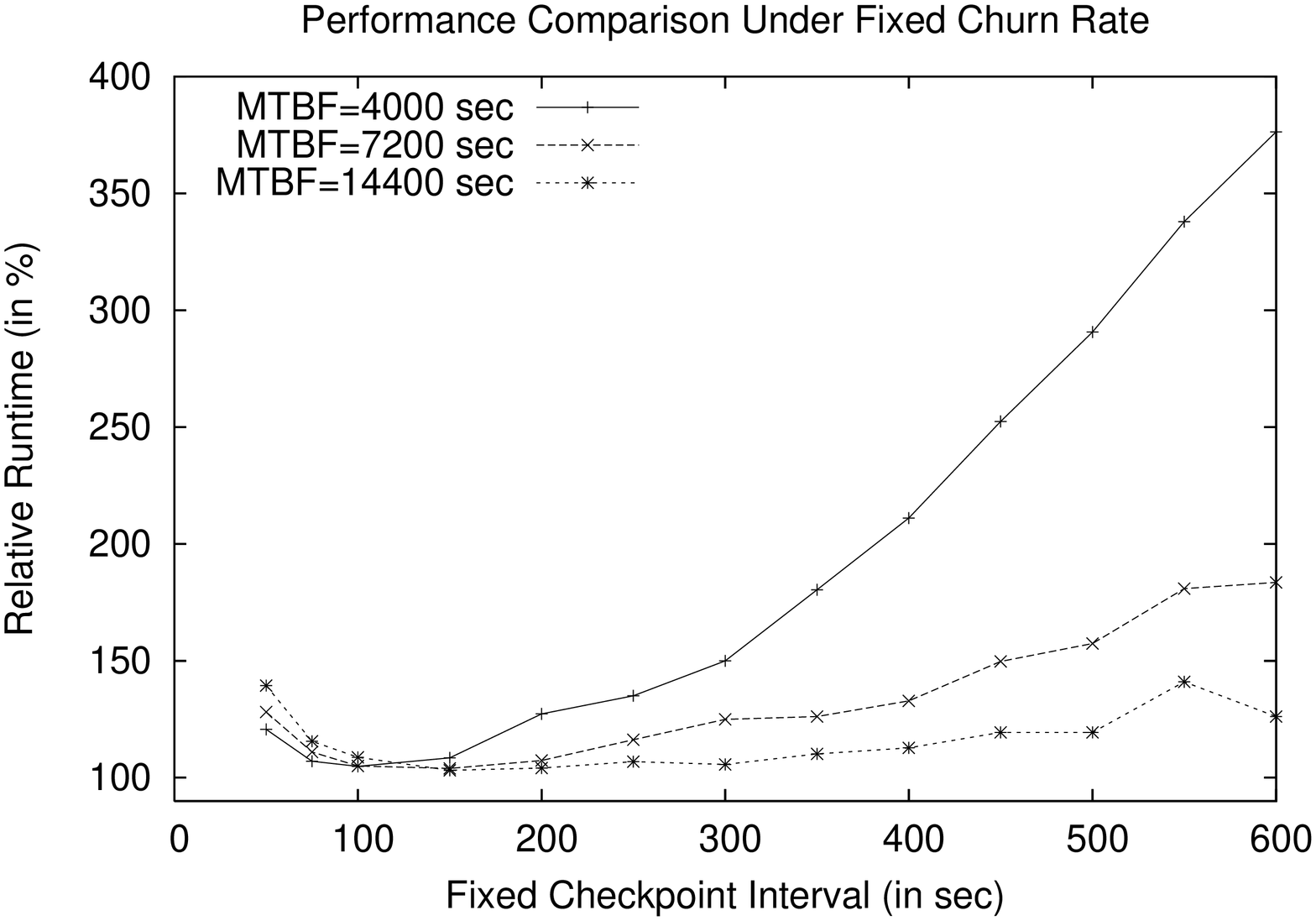}
\end{minipage}
\begin{minipage}[b]{0.33\linewidth}
\centering
\includegraphics[width=6.0cm, angle=270]{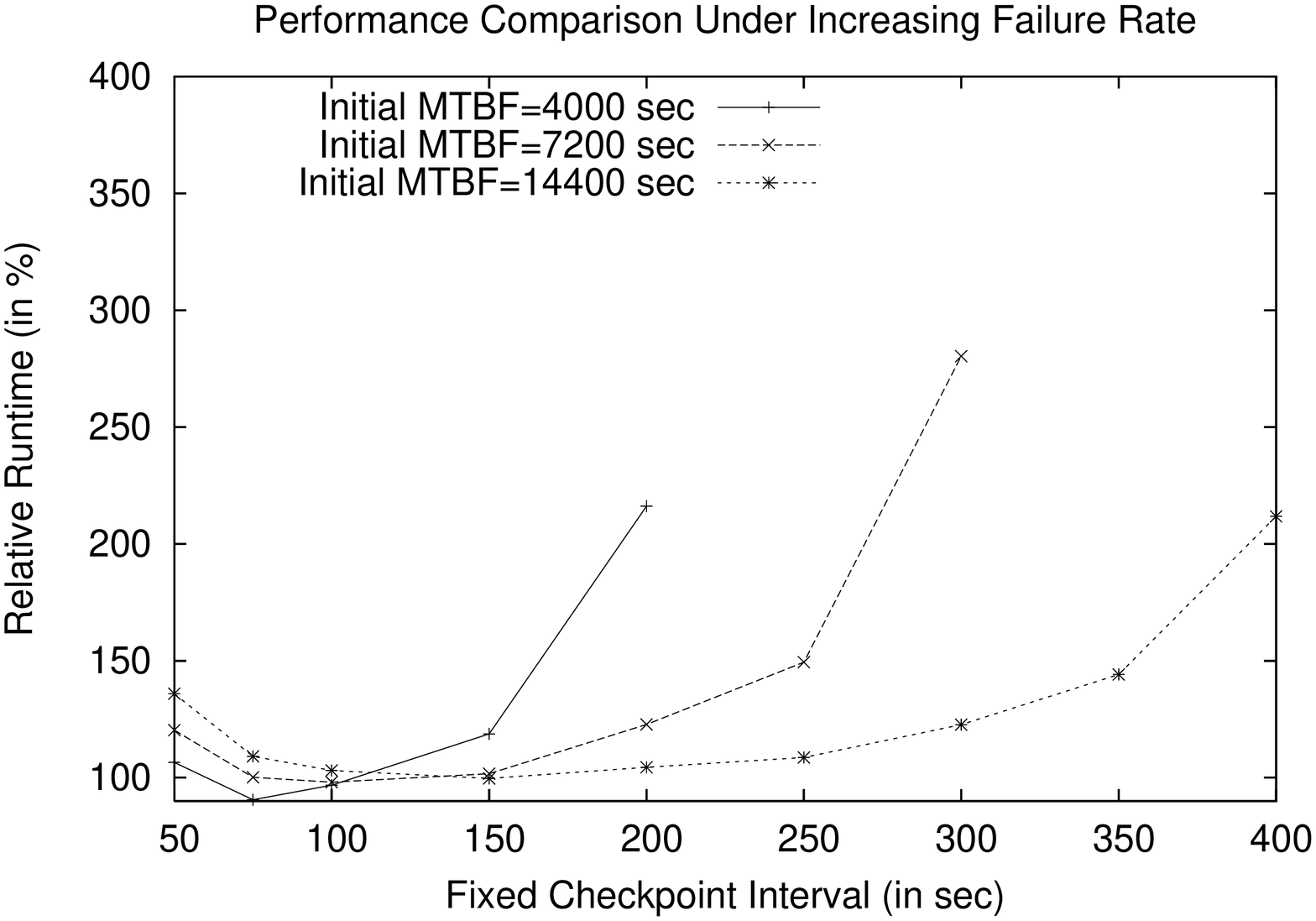}
\end{minipage}
\caption{Performance Comparison Between the Adaptive Checkpoint and Fixed Intervals Approach.}
\label{ckptperformance}
\end{figure*}

In \reffig{ckptperformance}, the performance of the proposed adaptive scheme is compared with the naive fixed
checkpoint interval approach under different departure rates. In this experiment, we set the checkpoint overhead 
to be 20 seconds and the image download overhead to be 50 seconds. The left chart of \reffig{ckptperformance} shows 
our adaptive scheme out performs the naive approach in all three (MTBF=4000, 7200 and 14400 seconds) different 
network departure rate environment which are typical settings to represent high, normal and low departure rates.
This is the failure rates set for the experiments and each peer would estimate the current peer failure rate, which 
would usually carry 10-15\% error as reported in~\cite{estimation}, for computing the optimized $\lambda$. 
The right chart of \reffig{ckptperformance} plots the results of experiments in which the departure rates are 
doubled in 20 hours with different initial departure rate. We choose the 20 hours failure rate double time as such
dynamism is observed in the Overnet trace data, all other settings are the same as the ones used in the previous experiment. 
As we can see, our scheme again out performed the naive approach in almost all cases and it helps to ensure
the jobs can all be eventually finished no matter of the network conditions while an arbitrarily selected checkpoint 
fixed checkpoint interval may cause the jobs running for ultra long time, for example, as we can see in the right 
chart of \reffig{ckptperformance}, compared with the adaptive checkpoint scheme, it took 3 times the runtime to finish 
the job when the initial nodes departure rate is MTBF=7200 seconds and the checkpoint interval set to be 5 minutes, it would 
take even much longer if the checkpoint interval is set to be even larger. The probability for the job to fail between 
two continuous checkpoints can increase rapidly as the checkpoint interval is too long and the job will keep rolling 
back to the same save status again and again. Such situation can be handled in our scheme as the checkpoint 
interval is always set according to the current estimated network conditions.    
    
\begin{figure*}
\begin{minipage}[b]{0.46\linewidth}
\centering
\includegraphics[width=6.0cm, angle=270]{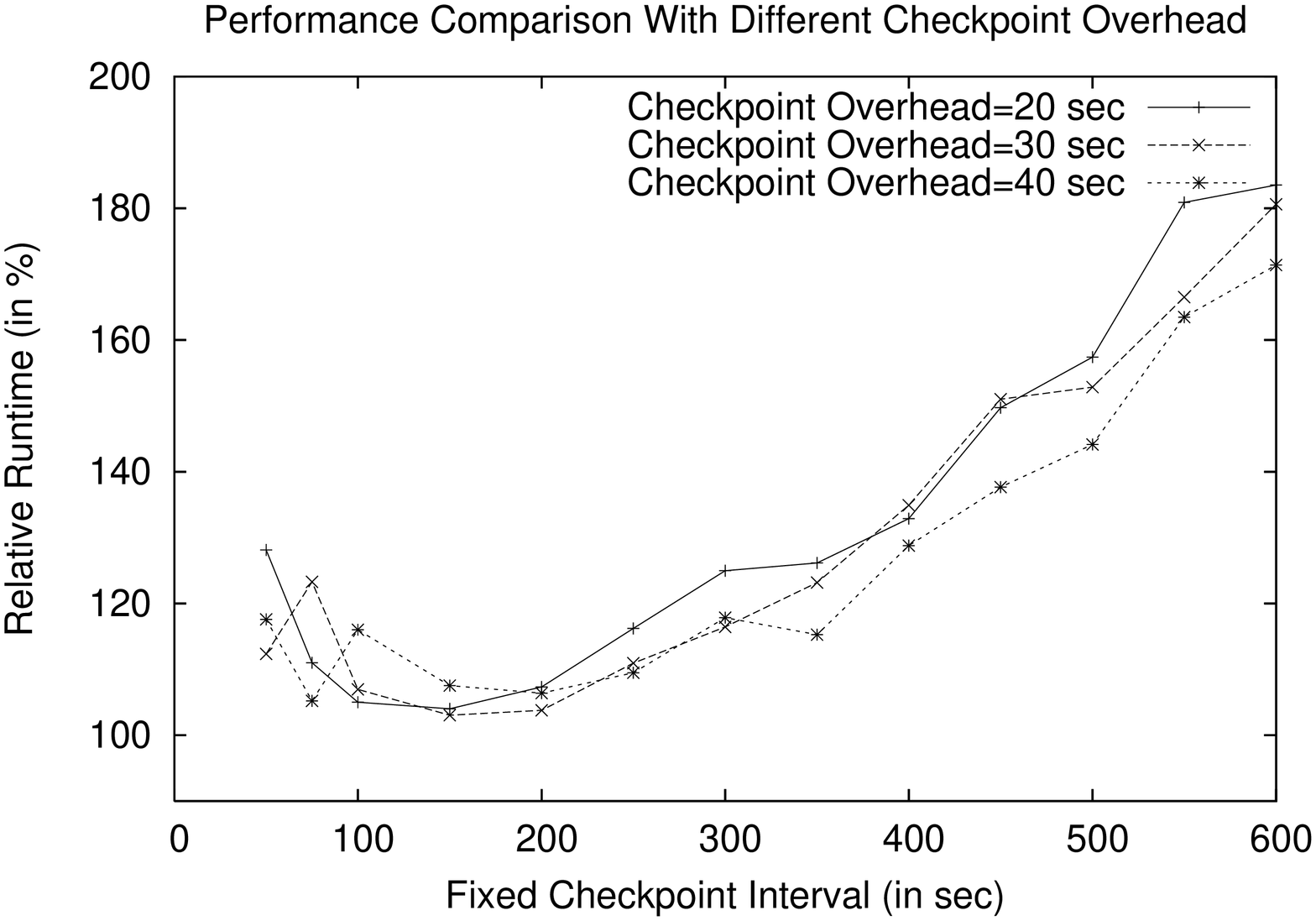}
\end{minipage}
\begin{minipage}[b]{0.33\linewidth}
\centering
\includegraphics[width=6.0cm, angle=270]{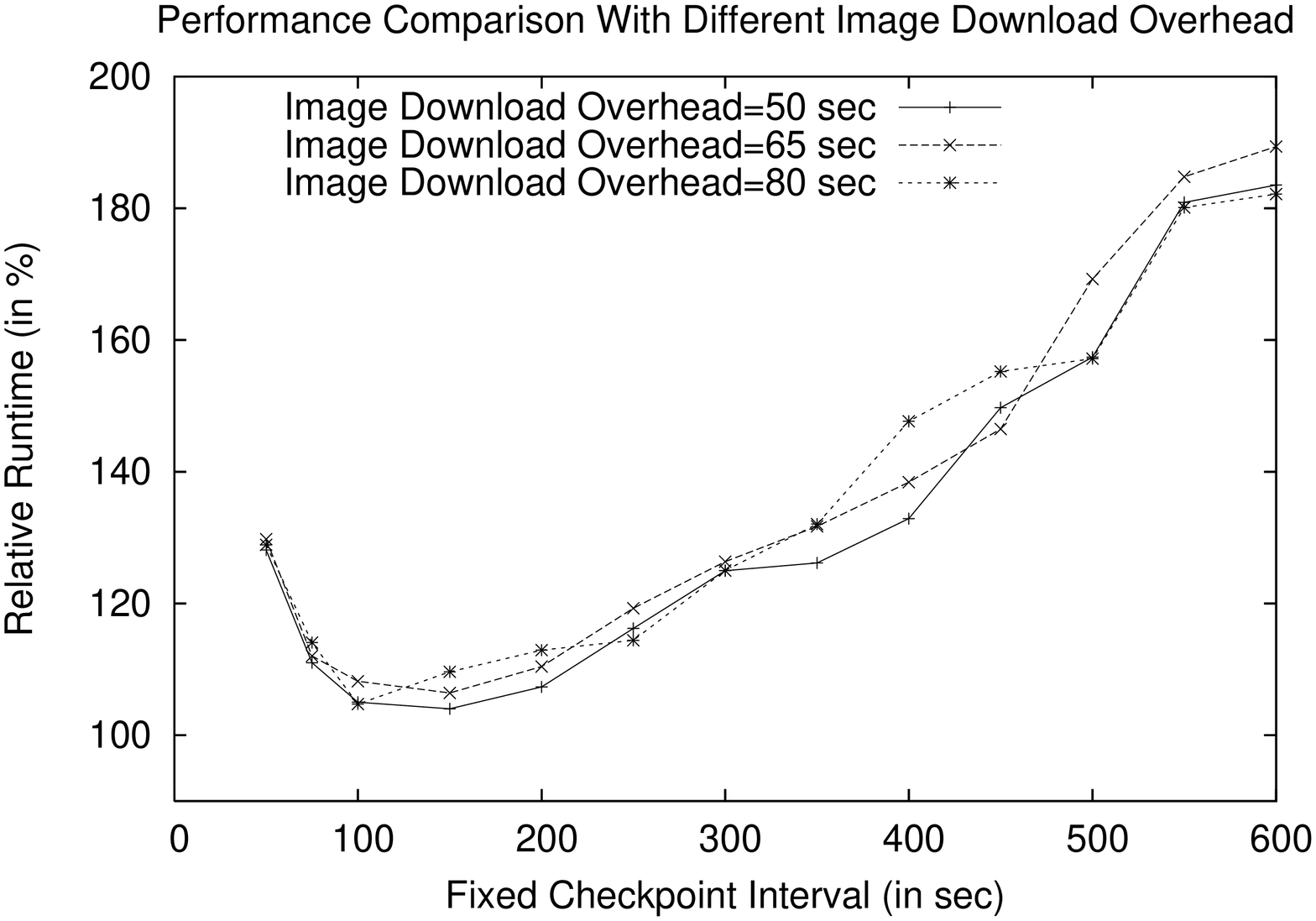}
\end{minipage}
\caption{Performance Comparison Under Different overheads.}
\label{varoverheads}
\end{figure*}

We also  simulated the performance of our scheme under different checkpoint overheads and image download overheads in 
\reffig{varoverheads}. As explained earlier, the checkpoint overhead means how long each checkpoint operation
can slow down the job, different checkpoint overhead can be caused due the types of programs as when the checkpoint 
images are being uploaded onto the network the bandwidth of upstream link of the desktops will be mostly used and 
the communication for message passing is going to become slowed down, thus programs in which processes need to 
communicate a lot with each other are going to suffer larger overheads. The checkpoint image download overhead is 
mainly determined by the available download bandwidth and can be approximated as the required time for the slowest 
node used in the job to download the checkpoint image. 

We first fixed the image download overhead at 50 seconds and tested our scheme with different checkpoint overheads.
The departure rate is MTBF=7200 seconds which represents the typical network condition according to the real P2P trace data
discussed earlier. The result is presented in the left chart in \reffig{varoverheads}. The right chart in 
\reffig{varoverheads} shows the results when we have fixed checkpoint overhead and various image download overheads. 
Our scheme demonstrated good adaptiveness in these two experiments and reported better performance compared to all 
tested fixed checkpoint intervals. 

\subsection{Discussion}
The experiment results of this work have shown a better optimized checkpoint and rollback approach can be achieved by 
applying the proposed adaptive checkpoint scheme to set the checkpoint interval; the performance of the running
message passing jobs can be better guarded by balancing the overheads caused by performing the checkpoints and
the overheads of actually restarting from a failure, which in turn is through monitoring the P2P network conditions.
The MTBF for each peer is only hours in three different 
deployed P2P networks, thus when several peers are used for a message passing processes, the MTBF for this group of peers is 
will be only around 5-10 minutes. We believe further measures can be taken to better handle highly dynamic networks. 
For example, we can potentially combine the process replication and checkpoint together to reduce the overheads of 
restarting. In such upgraded robustness design, jobs will only need to rollback to the previous known status only if all 
replicas of a process have failed, which can be less frequently and will increase the MTBF of the job.

%% file: ckpt_modeling.bbl
\begin{thebibliography}{10}

\bibitem{anderson04}
D.~P. Anderson.
\newblock Boinc: A system for public-resource computing and storage.
\newblock In {\em Proceedings of the 5th International Workshop on Grid
  Computing (GRID 2004)}, pages 4--10, 2004.

\bibitem{setiathome}
D.~P. Anderson, J.~Cobb, E.~Korpela, M.~Lebofsky, and D.~Werthimer.
\newblock Seti@home: an experiment in public-resource computing.
\newblock {\em Commun. ACM}, 45(11):56--61, 2002.

\bibitem{volunteercomputing}
D.~P. Anderson and G.~Fedak.
\newblock The computational and storage potential of volunteer computing.
\newblock In {\em CCGRID '06: Proceedings of the Sixth IEEE International
  Symposium on Cluster Computing and the Grid (CCGRID'06)}, pages 73--80,
  Washington, DC, USA, 2006. IEEE Computer Society.

\bibitem{BhagwanSV03}
R.~Bhagwan, S.~Savage, and G.~Voelker.
\newblock Understanding availability.
\newblock In {\em Proceedings of the 2nd International Workshop on Peer-to-Peer
  Systems (IPTPS '03)}, pages 256--267, Feb. 2003.

\bibitem{BHKLC06}
A.~Bouteiler, T.~Herault, G.~Krawezik, P.~Lemarinier, and F.~Cappello.
\newblock {MPICH-V} project: a multiprotocol automatic fault tolerant {MPI}.
\newblock {\em The International Journal of High Performance Computing
  Applications}, 20:319--333, Summer 2006.

\bibitem{mspastry}
M.~Castro, M.~Costa, and A.~Rowstron.
\newblock Performance and dependability of structured peer-to-peer overlays.
\newblock In {\em DSN '04: Proceedings of the 2004 International Conference on
  Dependable Systems and Networks (DSN'04)}, pages 9--18, Washington, DC, USA,
  2004. IEEE Computer Society.

\bibitem{ChandyL85}
K.~M. Chandy and L.~Lamport.
\newblock Distributed snapshots: Determining global states of distributed
  systems.
\newblock {\em ACM Transactions Computer Systems}, 3(1):63--75, 1985.

\bibitem{DBLP:journals/csur/ElnozahyAWJ02}
E.~N. Elnozahy, L.~Alvisi, Y.-M. Wang, and D.~B. Johnson.
\newblock A survey of rollback-recovery protocols in message-passing systems.
\newblock {\em ACM Computing Surveys}, 34(3):375--408, 2002.

\bibitem{p2pmpi}
S.~Genaud and C.~Rattanapoka.
\newblock A peer-to-peer framework for robust execution of message passing
  parallel programs.
\newblock In B.~D.~M. et~al., editor, {\em EuroPVM/MPI 2005}, volume 3666 of
  {\em LNCS}, pages 276--284. Springer-Verlag, September 2005.

\bibitem{GhinitaT06}
G.~Ghinita and Y.~M. Teo.
\newblock An adaptive stabilization framework for distributed hash tables.
\newblock In {\em 20th International Parallel and Distributed Processing
  Symposium (IPDPS 2006)}, Rhodes Island, Greece, 2006.

\bibitem{larson-2003}
S.~M. Larson, C.~D. Snow, M.~Shirts, and V.~S. Pande.
\newblock Folding@home and genome@home: Using distributed computing to tackle
  previously intractable problems in computational biology.
\newblock {\em Modern Methods in Computational Biology, Horizon Press}, 2003.

\bibitem{kademlia}
P.~Maymounkov and D.~Mazi{\`e}res.
\newblock Kademlia: A peer-to-peer information system based on the xor metric.
\newblock In {\em Proceedings of the 2nd International Workshop on Peer-to-Peer
  Systems (IPTPS '03)}, pages 53--65, 2002.

\bibitem{predictor}
J.~W. Mickens and B.~D. Noble.
\newblock Exploiting availability prediction in distributed systems.
\newblock In {\em NSDI'06: Proceedings of the 3rd conference on 3rd Symposium
  on Networked Systems Design \& Implementation}, pages 6--6, Berkeley, CA,
  USA, 2006. USENIX Association.

\bibitem{mpichopen}
L.~Ni and A.~Harwood.
\newblock An implementation of the message passing interface over an adaptive
  peer-to-peer network.
\newblock In {\em Proceedings of the 15th IEEE International Symposium on High
  Performance Distributed Computing (HPDC)}, pages 371--372, June 2006.

\bibitem{estimation}
L.~Ni and A.~Harwood.
\newblock A comparative study on peer-to-peer failure rate estimation.
\newblock In {\em Proceedings (To Appear) of the International Workshop on
  Peer-to-Peer Network Virtual Environments 2007, in conjunction with The 13th
  International Conference on Parallel and Distributed Systems.}, 2007.

\bibitem{dvm}
L.~Ni, A.~Harwood, and P.~J. Stuckey.
\newblock Realizing the e-science desktop peer using a peer-to-peer distributed
  virtual machine middleware.
\newblock In {\em MCG '06: Proceedings of the 4th international workshop on
  Middleware for grid computing}, page 7 pages, New York, NY, USA, 2006. ACM
  Press.

\bibitem{NurmiBW05}
D.~Nurmi, J.~Brevik, and R.~Wolski.
\newblock Modeling machine availability in enterprise and wide-area distributed
  computing environments.
\newblock In J.~C. Cunha and P.~D. Medeiros, editors, {\em Euro-Par}, volume
  3648 of {\em Lecture Notes in Computer Science}, pages 432--441. Springer,
  2005.

\bibitem{Powlese05BitTorrent}
J.~A. Pouwelse, P.~Garbacki, D.~H.~J. Epema, and H.~J. Sips.
\newblock The bittorrent p2p file-sharing system: Measurements and analysis.
\newblock In {\em Proceedings of the 4th International Workshop on Peer-to-Peer
  Systems (IPTPS '05)}, 2005.

\bibitem{935629}
L.~F.~G. Sarmenta.
\newblock {\em Volunteer computing}.
\newblock PhD thesis, 2001.
\newblock Supervisor-Stephen A. Ward.

\bibitem{chord}
I.~Stoica, R.~Morris, D.~Liben-Nowell, D.~R. Karger, M.~F. Kaashoek, F.~Dabek,
  and H.~Balakrishnan.
\newblock Chord: a scalable peer-to-peer lookup protocol for internet
  applications.
\newblock {\em IEEE/ACM Trans. Netw.}, 11(1):17--32, 2003.

\bibitem{gossip}
Th, R.~Guerraoui, S.~B. Handurukande, P.~Kouznetsov, and A.~M. Kermarrec.
\newblock Lightweight probabilistic broadcast.
\newblock {\em ACM Trans. Comput. Syst.}, 21(4):341--374, November 2003.

\bibitem{tianjing07}
J.~Tian and Y.~Dai.
\newblock Understanding the dynamic of peer-to-peer systems.
\newblock In {\em Proceedings of the 6th International Workshop on Peer-to-Peer
  Systems (IPTPS '07)}, Feb. 2007.

\end{thebibliography}
